\documentclass{moriond}

\usepackage{amsmath}

\def\dsttaunu{\bar{B} \rightarrow D^{*} \tau^- \bar{\nu}_\tau}
\def\ddsttaunu{\bar{B} \rightarrow D^{(*)} \tau^- \bar{\nu}_\tau}

\def\dstlnu{\bar{B} \rightarrow D^{*} \ell^- \bar{\nu}_\ell}
\def\ddstlnu{\bar{B} \rightarrow D^{(*)} \ell^- \bar{\nu}_\ell}
\def\taunu{B^- \rightarrow \tau^- \bar{\nu}_\tau}
\def\Bsig{B_{\rm sig}}
\def\Btag{B_{\rm tag}}
\def\pinu{\tau^- \rightarrow \pi^- \nu_\tau}
\def\rhonu{\tau^- \rightarrow \rho^- \nu_\tau}
\def\lnunu{\tau^- \rightarrow \ell^- \bar{\nu}_\ell \nu_\tau}

\newcommand{\Photo}{}

\begin{document}
\vspace*{4cm}
\title{$\ddsttaunu$ and Related Tauonic Topics at Belle}

\author{S.~Hirose, For the Belle Collaboration}

\address{KMI, Nagoya University, Furo, Chikusa, Nagoya, Japan}

\maketitle\abstracts{
  The decays $\ddsttaunu$ are good probes to new physics beyond the Standard Model. The ratios of branching fractions $R(D^{(*)}) \equiv \mathit{BF}(\ddsttaunu) / \mathit{BF}(\ddstlnu)$ (where $\ell^- = e^-, \mu^-$) measured by Belle, BaBar and LHCb show 3.9$\sigma$ deviation from the SM expectations as of 2015. In 2016, the Belle collaboration has shown two new measurements for the $\ddsttaunu$ decay. These include the first application of the semileptonic tagging to the $R(D^*)$ measurement and the first measurement of the $\tau$ polarization using the hadronic $\tau$ decays. We also review the two measurements for $\taunu$ at Belle. Along with these results, compatibility with the type-II Two-Higgs-Doublet Model is discussed. 
}

\section{Introduction}

Semileptonic and leptonic $B$ meson decays containing a $\tau$ lepton in the final state are theoretically well-understood processes in the Standard Model (SM)~\cite{cite:Hwang:2000}. Owing to the existence of the two heavy fermions, a $b$ quark and a $\tau$ lepton, they are sensitive to new physics (NP) beyond the SM if the NP has an enhanced coupling to the third-generation fermions. The decays $\ddsttaunu$ and $\taunu$ are these types of the $B$ decays which have been experimentally investigated by the $B$-factory experiments, Belle and BaBar.\footnote{Throughout this paper, the inclusion of the charge-conjugate mode is always implied.} LHCb has also demonstrated their capability of studying the $\dsttaunu$ process at the Large Hadron Collider.

In this paper, we discuss the recent experimental results on $\ddsttaunu$ and $\taunu$ at the Belle experiment, where 8~GeV electrons and 3.5~GeV positrons are collided by the KEKB accelerator at the center-of-mass energy of 10.58~GeV. Pairs of $B$ mesons are produced through the process $\Upsilon(4S) \rightarrow B \bar{B}$. The $B$ meson decays are recorded by the Belle detector, which is a complex of six sub-detector systems~\cite{cite:Abashian:2002}.

\section{Tagging Method}

Due to multiple neutrinos in the final state, signal $B$ mesons ($\Bsig$) decaying into $D^{(*)} \tau^- \bar{\nu}_\tau$ or $\tau^- \bar{\nu}_\tau$ cannot be fully reconstructed. Exploiting the advantage of Belle that no particle except for two $B$ mesons is produced in the $\Upsilon(4S)$ decay, we ``tag'' the $\Bsig$ candidate by reconstructing the counterpart $B$ meson ($\Btag$) at first. The remaining particles in the event are then assigned to form a $\Bsig$ candidate.

There are two tagging methods applied to the $\ddsttaunu$ and $\taunu$ analyses. In the hadronic tagging method, $\Btag$ is fully reconstructed from one of the hadronic decay modes. The four-momentum of $\Bsig$ is extracted by $p_{\Bsig} = p_{e^+ e^-} - p_{\Btag}$, where $p$'s are the four-momenta of the $\Bsig$ the $e^+ e^-$ beam and the $\Btag$, respectively. Belle has developed a hadronic tagging algorithm based on the NeuroBayes neural-network package~\cite{cite:Feindt:2011}. This algorithm uses 1104 decay chains in total, to achieve the highest possible $\Btag$ reconstruction efficiency. The typical efficiency is around 0.2--0.3\%.

The semileptonic tagging method uses the semileptonic decays such as $\ddstlnu$. Although one neutrino in the semileptonic decay makes full reconstruction of $\Btag$ impossible, the $\Btag$ candidates are identified using the variable 
\begin{eqnarray}
  \cos\theta_{B - D^{(*)} \ell} = \frac{2 E_{\rm beam}^* E_{D^{(*)} \ell}^* - m_B^2c^4 - M_{D^{(*)} \ell}^2c^4}{2 |\vec{p}_B^{\kern2pt *}| |\vec{p}_{D^{(*)} \ell}^{\kern2pt *}| c^2},
\end{eqnarray}
where $E^*$, $\vec{p}^{\kern2pt *}$ and $m$ ($M$) denote the energy, the momentum and the (reconstructed) mass, respectively, with the subscripts representing the $e^+ e^-$ beam, the $\Btag$ and the $D^{(*)} \ell$ system. By requiring $\cos \theta_{B - D^{(*)} \ell}$ to lie in the physical region between $-1$ and $1$, correct $\Btag$ candidates are obtained.

\section{Experimental Results from Belle}

\subsection{Situation of $\ddsttaunu$ Studies before Winter 2016}

The decays $\ddsttaunu$ have the relatively large branching fraction of $\mathcal{O}(1)$\% among the $B$ meson decay modes. Its three-body decay realizes to probe NP amplitudes using its kinematics such as the $\tau$ polarization, not only the branching fraction. To study the decays $\ddsttaunu$, the ratios of the branching fractions
\begin{eqnarray}
  R(D^{(*)}) &\equiv& \frac{\mathit{BF}(\bar{B} \rightarrow D^{(*)} \tau^- \bar{\nu}_\tau)}{\mathit{BF}(\bar{B} \rightarrow D^{(*)} \ell^- \bar{\nu}_\ell)}
\end{eqnarray}
are measured, where $\ell^-$ is an electron or a muon. In the ratios, the uncertainties in the Cabibbo-Kobayashi-Maskawa matrix element $|V_{cb}|$, the hadronic form factors and the experimental reconstruction efficiency are largely canceled. The SM predicts $R(D) = 0.300 \pm 0.008$~\cite{cite:HPQCD:2015} and $R(D^*) = 0.252 \pm 0.003$~\cite{cite:Fajfer:2012}.

With the full data sample, Belle performed a measurement of $R(D^{(*)})$ using the hadronic tagging and the leptonic final state of the $\tau$ lepton~\cite{cite:Belle:2015}. The result was compatible with the SM expectation within 1.8$\sigma$. Including this result and the results from BaBar~\cite{cite:BaBar:2012} and LHCb~\cite{cite:LHCb:2015}, the world-average $R(D)$ and $R(D^*)$ estimated by the heavy-flavor-averaging group (HFAG) were $0.391 \pm 0.041({\rm stat.}) \pm 0.028({\rm syst.})$ and $0.322 \pm 0.018({\rm stat.}) \pm 0.012({\rm syst.})$, respectively~\cite{cite:HFAG:2016}. These were by 1.7 and 3.0$\sigma$ away from the expectation based on the SM. The overall discrepancy reached 3.9$\sigma$. 

\subsection{New Measurements for $\dsttaunu$}

In 2016, Belle has shown the second and the third measurements for $\dsttaunu$. The second one~\cite{cite:Belle:2016} is based on the semileptonic tagging and provides an independent $R(D^*)$ measurement from the previous study. Due to less constraints of the semileptonic tagging on the $\Btag$ kinematics, more background than the measurement with the hadronic tagging was predicted. Therefore only $R(D^*)$ has been measured from the $\bar{B}^0 \rightarrow D^{*+} \tau^- \bar{\nu}_\tau$ channel. For the $\Btag$ decay, $\bar{B}^0 \rightarrow D^{*+} \ell^- \bar{\nu}_\ell$ has been chosen. Namely, the $\dsttaunu$ (signal) event has $\bar{B}_{\rm sig}^0 \rightarrow D^{*+} \tau^- \bar{\nu}_\tau$ and $B_{\rm tag}^0 \rightarrow D^{*-} \ell^+ \nu_\ell$ while the $\dstlnu$ (normalization) event has $\bar{B}_{\rm sig}^0 \rightarrow D^{*+} \ell^- \bar{\nu}_\ell$ and $B_{\rm tag}^0 \rightarrow D^{*-} \ell^+ \nu_\ell$.

\begin{figure}[t!]
  \centering
  \includegraphics[width=8cm]{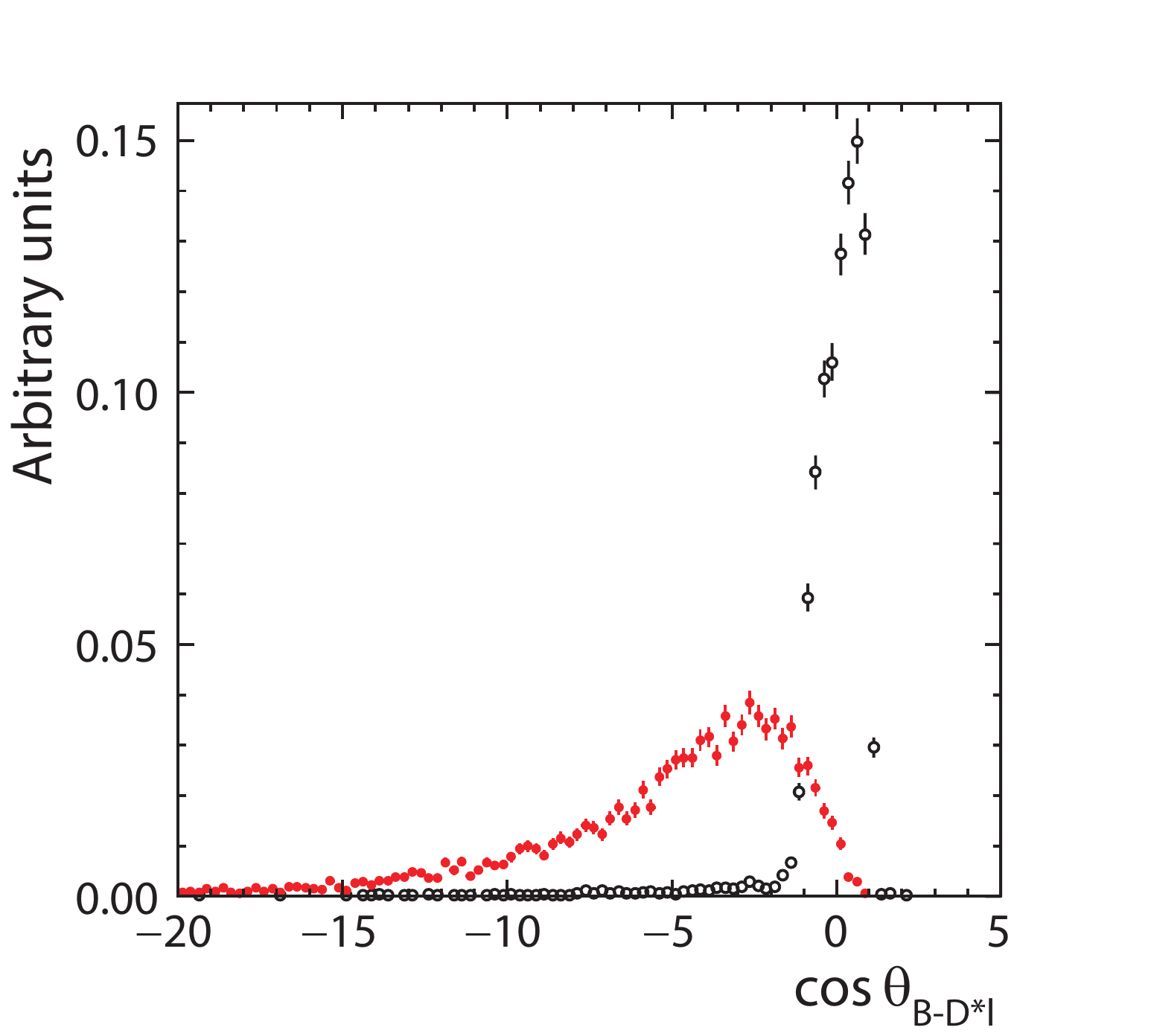}
  \caption{Distribution of $\cos\theta_{B - D^* \ell}^{\rm sig}$ for the signal events (solid red circles) and the normalization events (open black circles) from the Monte Carlo simulation.}
  \label{fig:cos_bdstl}
\end{figure}

In this measurement, both signal and normalization events have the same final state: two $D^*$, two $\ell$ and a missing momentum. It is therefore important to consider how to separate the signal events from the normalization events. In one event, two values of $\cos\theta_{B-D^* \ell}$ are defined for each of the two $B$ mesons. Due to three neutrinos, very often $\Bsig$ has $\cos\theta_{B-D^*\ell}$ significantly smaller than $-1$. The smaller value of the two $\cos \theta_{B - D^*\ell}$ ($\cos \theta_{B - D^* \ell}^{\rm sig}$) therefore provides efficient separation between the signal and the normalization events, as shown in Fig.~\ref{fig:cos_bdstl}~\cite{cite:Belle:2016}. Adding two more variables, a multi-variate analysis based on NeuroBayes is performed and the output classifier $O_{\rm NB}$ is constructed. Further details of the analysis are discussed in Ref.~\cite{cite:Belle:2016}.

\begin{figure}[t!]
  \centering
  \includegraphics[width=7.5cm]{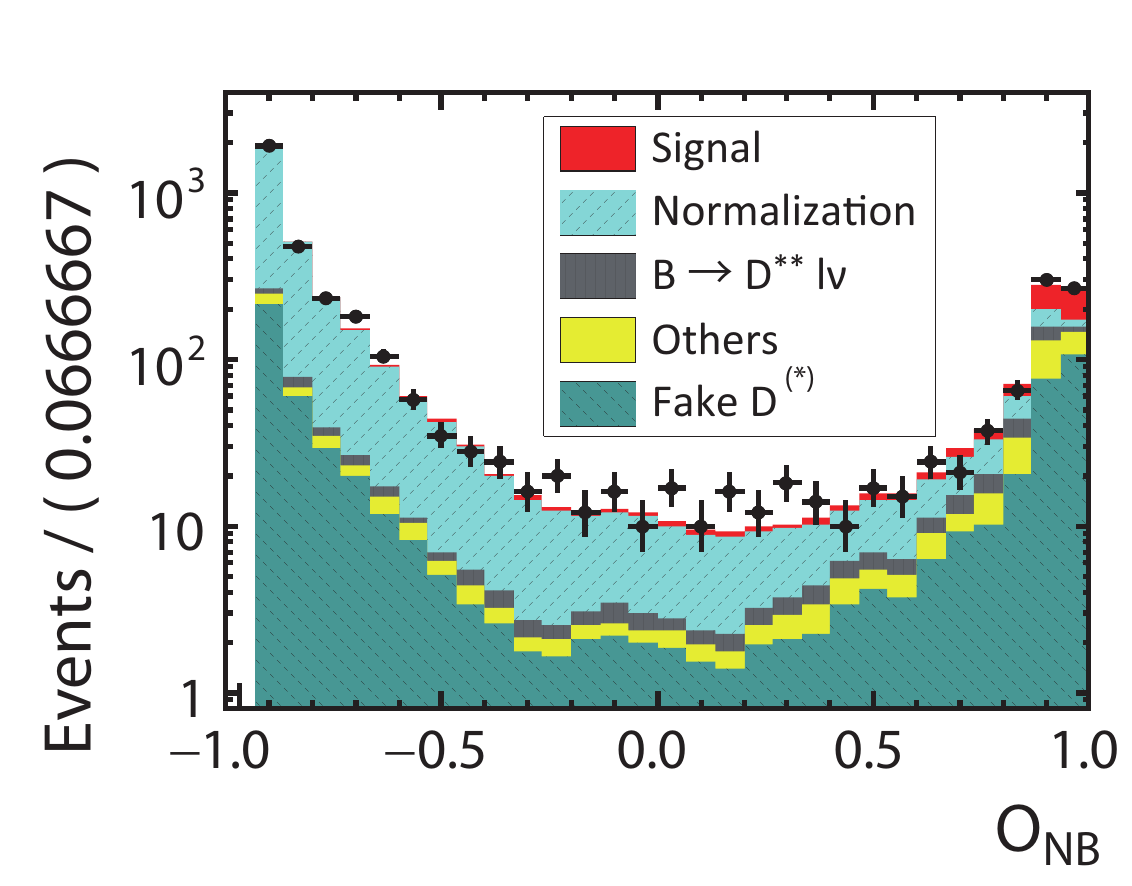}
  \includegraphics[width=7.5cm]{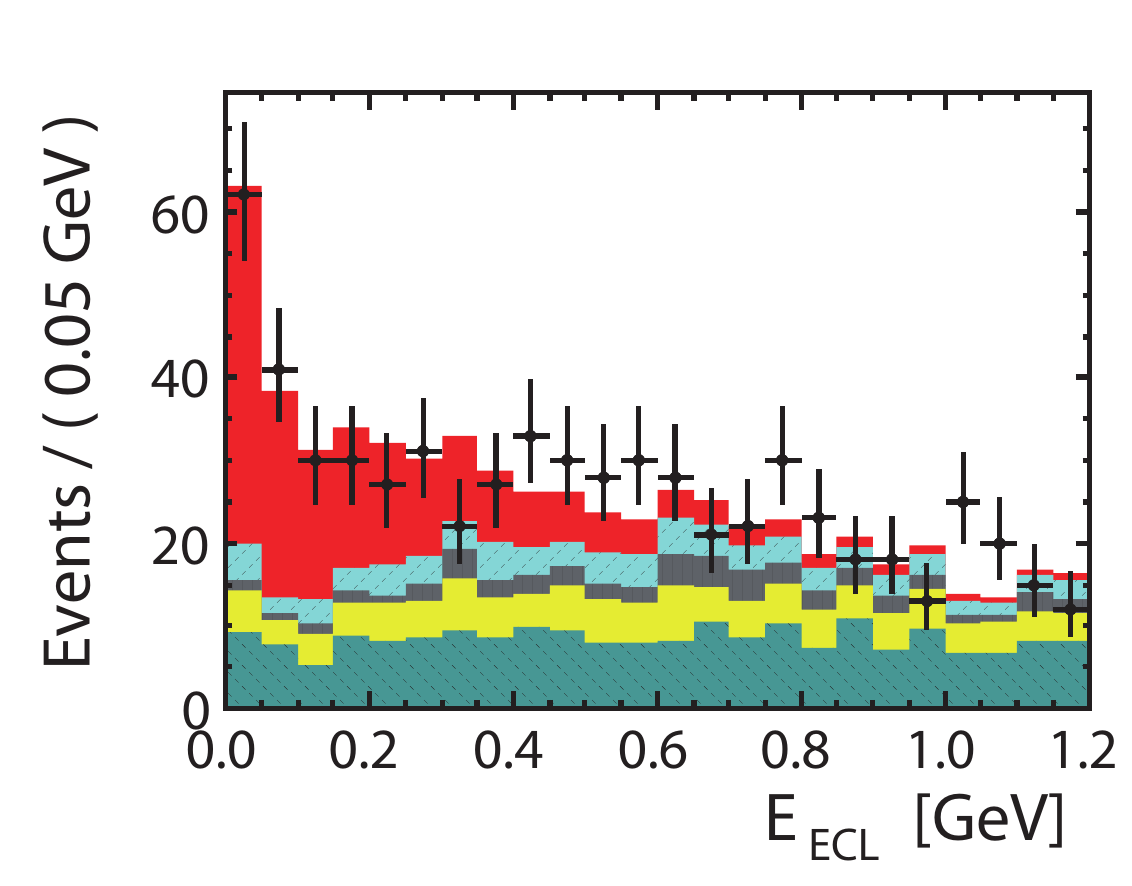}
  \caption{Two-dimensional fit result. (left) Projection to $O_{\rm NB}$. (right) Projection to $E_{\rm ECL}$ in the region $O_{\rm NB} > 0.8$. For both panels, the black dots show the distribution from the experimental data while the solid colored histograms are the fitted probability density functions constructed from the Monte-Carlo simulation sample.}
  \label{fig:sl-result}
\end{figure}

Figure~\ref{fig:sl-result} shows the result of the two-dimensional fit using $O_{\rm NB}$ and $E_{\rm ECL}$~\cite{cite:Belle:2016}. The second variable $E_{\rm ECL}$ is the energy sum of the clusters in the electromagnetic calorimeter that are not used for the event reconstruction. Compared to the signal and the normalization events, other background events tend to have larger values of $E_{\rm ECL}$ due to additional photons from $B$ meson decays. This measurement results in
\begin{eqnarray}
  R(D^*) &=& 0.302 \pm 0.030({\rm stat.}) \pm 0.011({\rm syst.}),
\end{eqnarray}
which is consistent with the SM within 1.6$\sigma$.

The third measurement~\cite{cite:Belle:2017} is based on the hadronic tagging and the hadronic $\tau$ decays $\pinu$ and $\rhonu$. This choice of the $\tau$ final states allows a new $R(D^*)$ measurement independent of the result with $\lnunu$. Since the $\Bsig$ final state contains only hadrons, the main background component in this study arises from hadronic $B$ decays. Their high-multiplicity final states through complicated hadronization processes make both experimental measurements and theoretical predictions difficult. Estimation of the amount of hadronic $B$ events in the signal region is thus one of the challenges. At the same time, this is an advantage as the background composition is different from the previous studies, where semileptonic $B$ decays with the excited $D$ mesons heavier than $D^*$ are one of the major sources of the systematic uncertainty.

In addition to the new measurement of $R(D^*)$, the two-body $\tau$ decays allow a measurement of the $\tau$ polarization. It is defined by
\begin{eqnarray}
  P_\tau(D^*) &\equiv& \frac{\Gamma^+(D^*) - \Gamma^-(D^*)} {\Gamma^+(D^*) + \Gamma^-(D^*)},
\end{eqnarray}
where $\Gamma^{+(-)}(D^*)$ is the decay rate for the $\tau$ lepton with a positive (negative) helicity state. The SM predicts $P_\tau(D^*) = -0.497 \pm 0.013$~\cite{cite:Tanaka:2013}. This quantity is experimentally extracted from the differential decay rate
\begin{eqnarray}
  \frac{1}{\Gamma(D^*)}\frac{d\Gamma(D^*)}{d \cos\theta_{\rm hel}} &=& \frac{1}{2}[1 + \alpha P_\tau(D^*) \cos\theta_{\rm hel}],\label{eq:coshel}
\end{eqnarray}
where $\Gamma(D^*)$ and $\theta_{\rm hel}$ denote the total decay rate and the angle of the $\tau$-daughter meson momentum with respect to the direction opposite the virtual $W$ boson\footnote{There are two virtual $W$ bosons in the $\bar{B} \rightarrow D^* \tau^- \bar{\nu}_\tau$ decay: one from the $B$ meson decay and the other from the $\tau$ lepton decay. In this paper, $W$ always denotes the virtual $W$ boson from the $B$ decay.} momentum in the rest frame of $\tau$. The coefficient $\alpha$ is represented by
\begin{eqnarray}
  \alpha &=&
  \begin{cases}
    1 & \text{for $\pinu$}\\
    \frac{m_{\tau}^2 - 2m_\rho^2}{m_{\tau}^2 + 2m_\rho^2} & \text{for $\rhonu$,}\label{eq-alpha}
  \end{cases}
\end{eqnarray}
where $m_\tau$ and $m_\rho$ are the masses of the $\tau$ lepton and the $\rho$ meson, respectively.

Based on Eq.~\ref{eq:coshel}, $P_\tau(D^*)$ is related to the forward-backward asymmetry of the signal distribution:
\begin{eqnarray}
  P_\tau(D^*) &=& \frac{2}{\alpha} \frac{N_{\rm sig}^{\rm F} - N_{\rm sig}^{\rm B}}{N_{\rm sig}^{\rm F} + N_{\rm sig}^{\rm B}},
\end{eqnarray}
where $N_{\rm sig}^{\rm F(B)}$ denotes the number of signal events in the region $\cos \theta_{\rm hel} > (<)~0$.

\begin{figure}[b!]
  \centering
  \includegraphics[width=13cm]{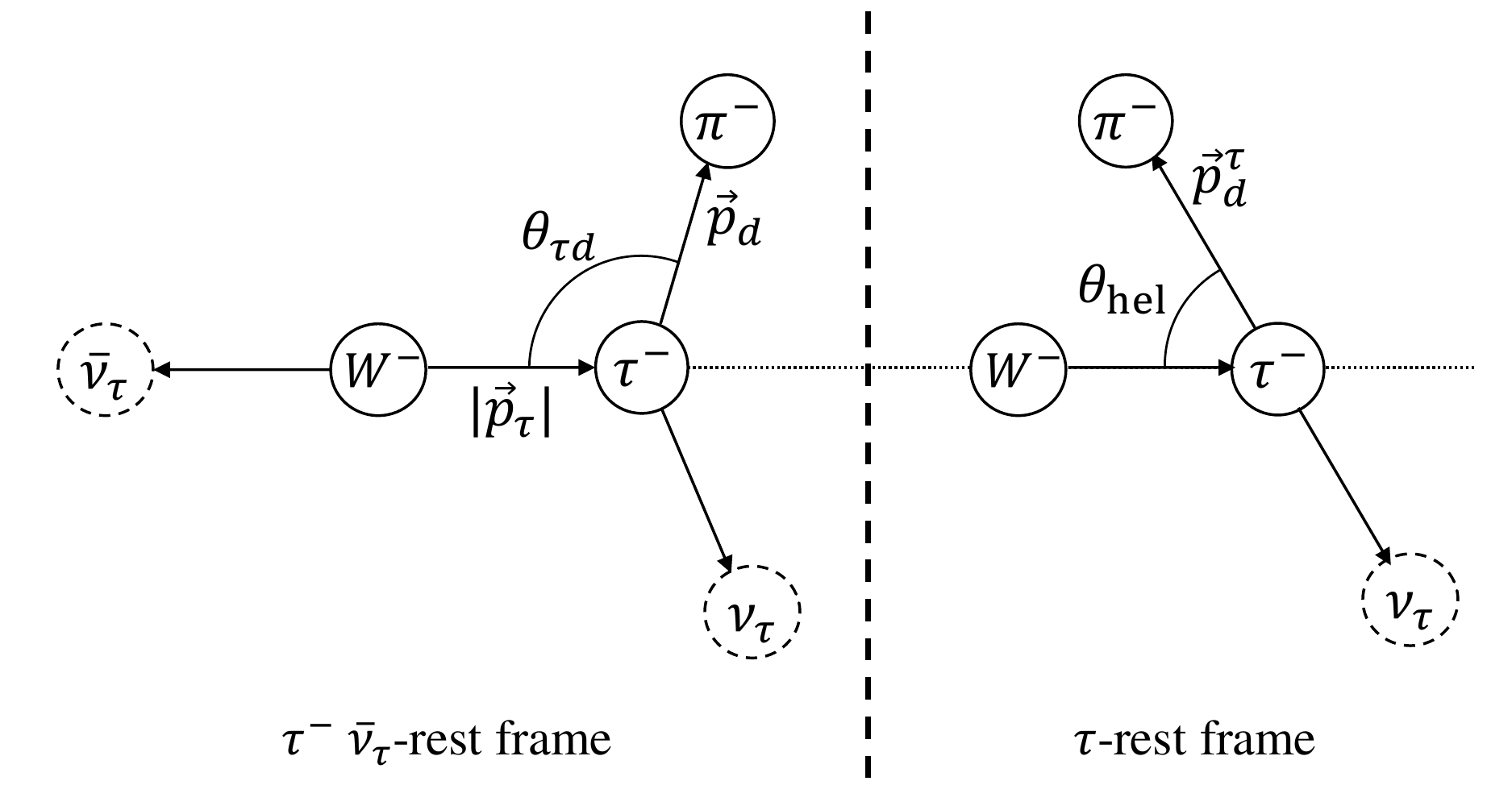}
  \caption{Kinematics in the rest frame of the $\tau^- \bar{\nu}_\tau$ system and in the rest frame of $\tau$.}
  \label{fig:frame}
\end{figure}

Experimentally, the rest frame of $\tau$ cannot be exactly taken since the $\tau$ momentum is not completely determined. We instead use the rest frame of the $\tau \bar{\nu}_\tau$ system. In this frame shown in Fig.~\ref{fig:frame}, the energy and the $\tau$ momentum are, respectively, determined by
\begin{eqnarray}
  E_\tau           &=& \frac{q^2 + m_{\tau}^2/c^2}{2 \sqrt{q^2}},\\
  |\vec{p}_{\tau}| &=& \frac{q^2 - m_{\tau}^2/c^2}{2 \sqrt{q^2}},
\end{eqnarray}
where $q^2 = p_{e^+ e^-} - p_{\Btag} - p_{D^*}$ and $p_{D^*}$ is the reconstructed four-momentum of $D^*$. Using $E_\tau$ and $|\vec{p}_\tau|$,
\begin{eqnarray}
  \cos \theta_{\tau d} &=& \frac{2 E_\tau E_d - m_{\tau}^2 c^4 - m_d^2 c^4} {2 |\vec{p}_{\tau}||\vec{p}_d| c^2},
\end{eqnarray}
is calculated, where $E$ and $\vec{p}$ are the energy and the three-momentum, respectively, of the $\tau$ lepton and the $\tau$-daughter meson $d = \pi$ or $\rho$. Using the Lorentz transformation from this frame to the rest frame of $\tau$, we obtain the equation
\begin{eqnarray}
  |\vec{p}_d^{\kern2pt \tau}| \cos\theta_{\rm hel} &=& -\gamma |\vec{\beta}| E_d / c + \gamma |\vec{p}_d| \cos\theta_{\tau d},\label{eq:Lorentz}
\end{eqnarray}
where
\begin{eqnarray}
  \gamma &=& \frac{E_\tau} {(m_\tau c^2)},\\
  |\vec{\beta}| &=& \frac{|\vec{p}_\tau|} {E_\tau}.
\end{eqnarray}
The $\tau$-daughter momentum in the rest frame of $\tau$ is represented by
\begin{eqnarray}
  |\vec{p}_d^{\kern2pt \tau}| &=& \frac{m_\tau^2 - m_d^2} {2 m_\tau}.
\end{eqnarray}
Solving Eq.~\ref{eq:Lorentz}, $\cos \theta_{\rm hel}$ is obtained.

\begin{figure}[b!]
  \centering
  \includegraphics[width=7cm]{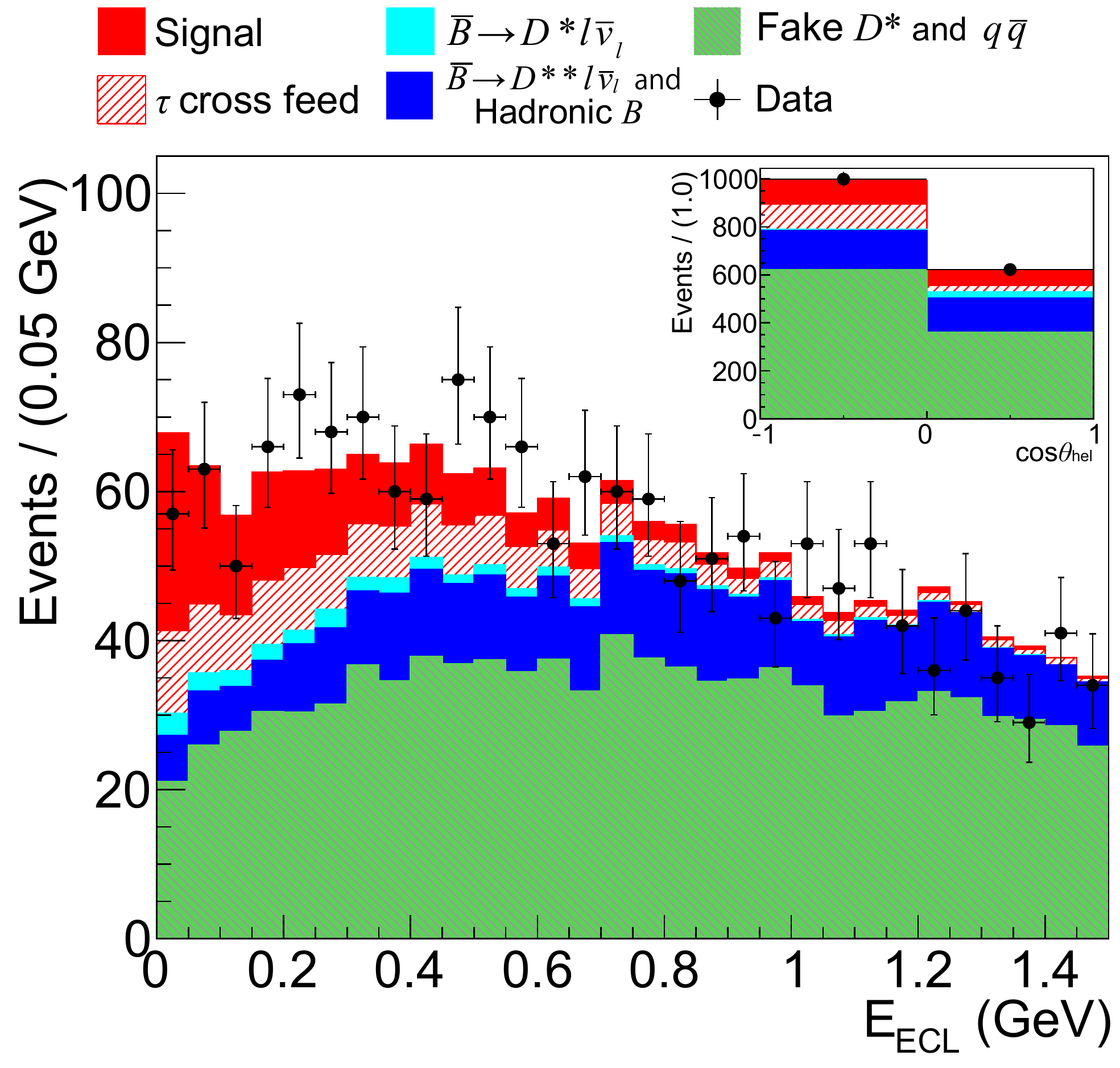}
  \includegraphics[width=8cm]{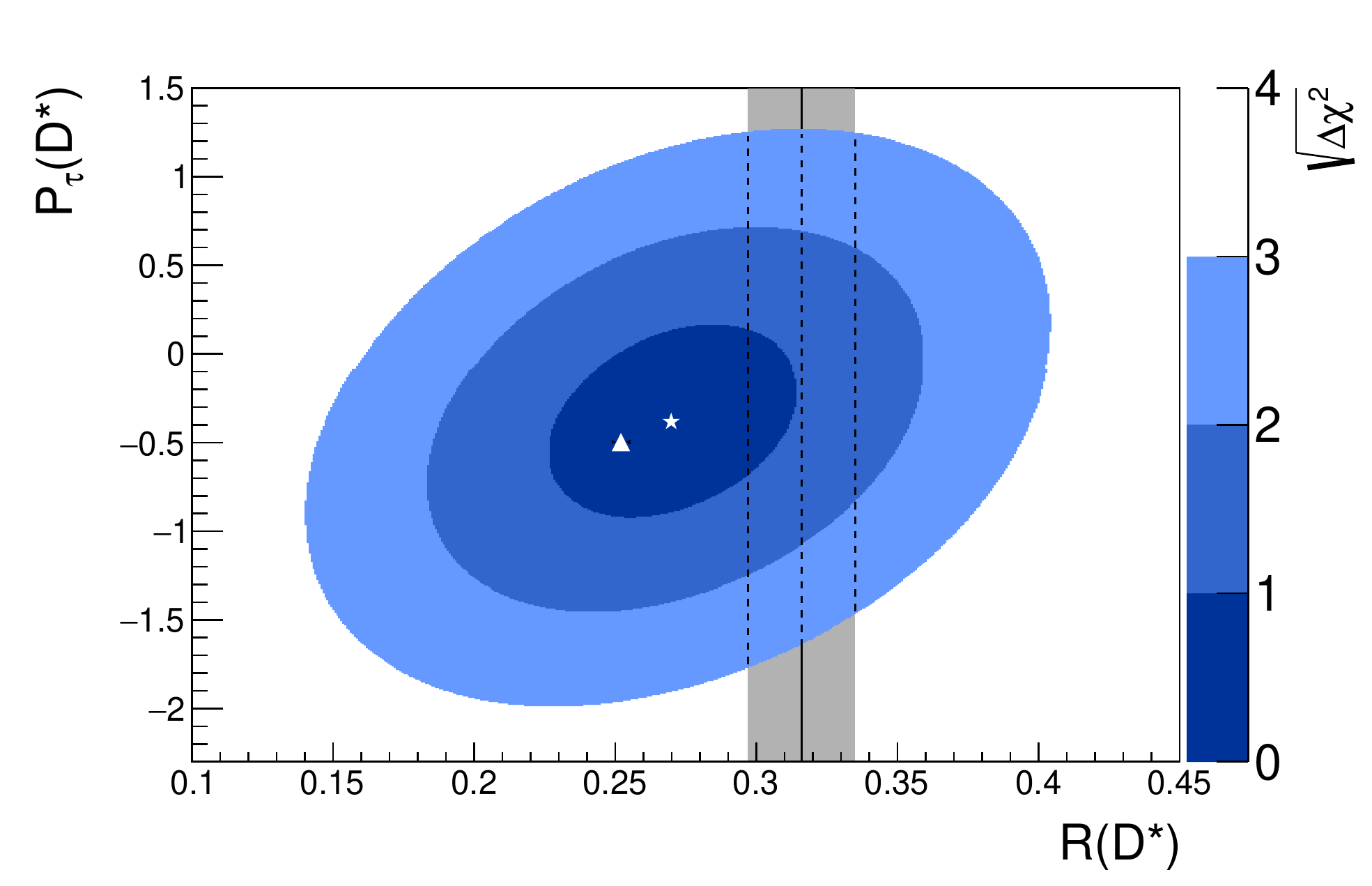}
  \caption{Result from the $R(D^*)$ and $P_\tau(D^*)$ measurement using hadronic $\tau$ decays. (left) Fit result projected to the $E_{\rm ECL}$ and $\cos\theta_{\rm hel}$ axes. (right) Comparison of our result (star for the best-fit value and 1$\sigma$, 2$\sigma$, 3$\sigma$ contours) with the SM prediction (triangle). The shaded vertical band shows the world average without this result.}
  \label{fig:rdspt}
\end{figure}

\begin{figure}[b!]
  \centering
  \includegraphics[width=10cm]{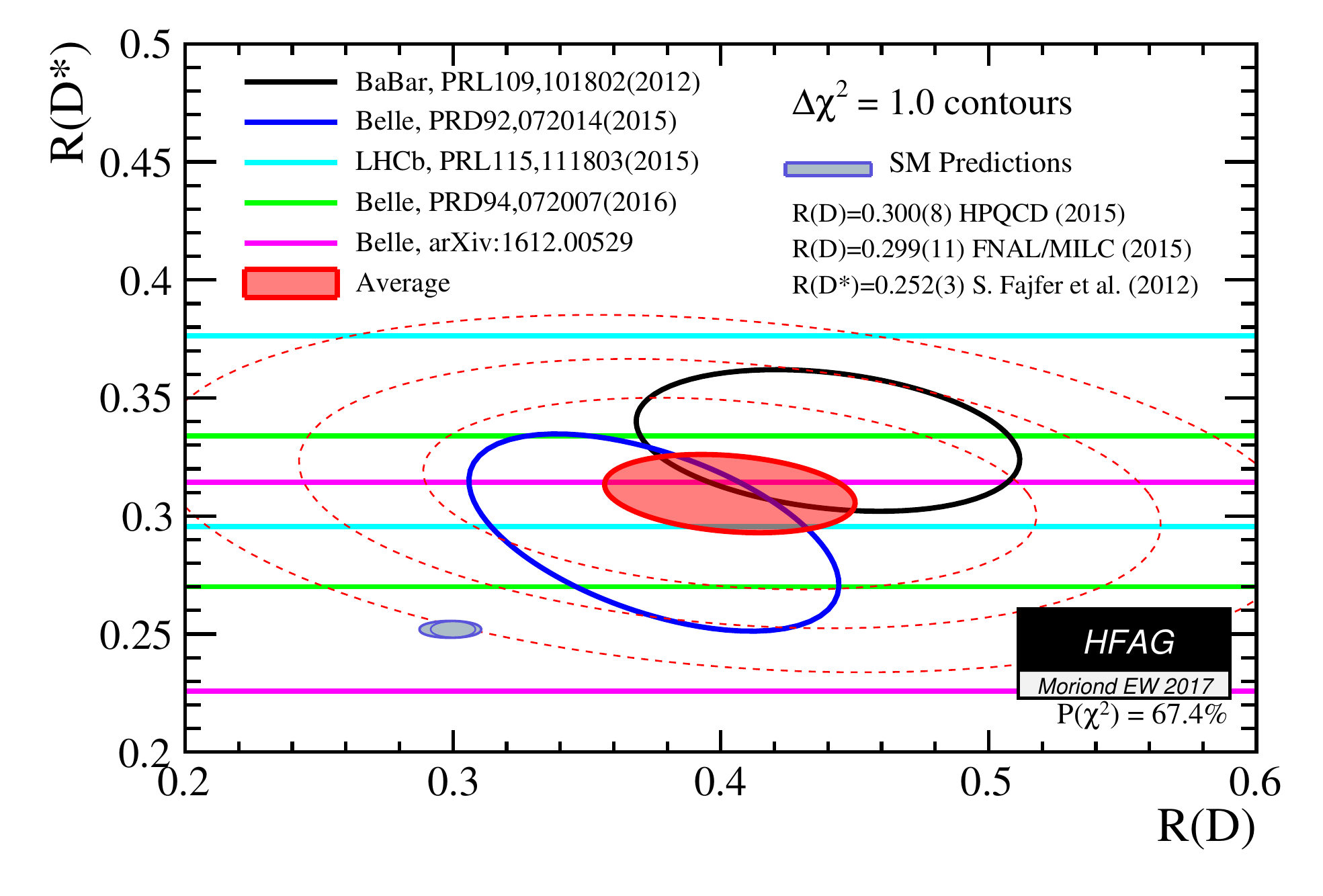}
  \caption{Comparison of the experimental results on $R(D^{(*)})$ with the SM expectations.}
  \label{fig:ddsttaunu_summary}
\end{figure}

A fit is performed in two steps. First, the yield of the normalization events is measured using the missing-mass squared
\begin{eqnarray}
    M_{\rm miss}^2 &=& (p_{e^+ e^-} - p_{\rm tag} - p_{D^*} - p_{\ell})^2 / c^2,
\end{eqnarray}
where $p_\ell$ denotes the four-momentum of $\ell$ and the other variables are defined previously. After determining the normalization yield, a two-dimensional fit is done using $E_{\rm ECL}$ and $\cos\theta_{\rm hel}$, as shown in Fig.~\ref{fig:rdspt} (left)~\cite{cite:Belle:2017}. The measurement results in
\begin{eqnarray}
  R(D^*) &=& 0.270 \pm 0.035({\rm stat.}) ^{+0.028}_{-0.025}({\rm syst.}),\\
  P_\tau(D^*) &=& -0.38 \pm 0.51({\rm stat.}) ^{+0.21}_{-0.61}({\rm syst.}).
\end{eqnarray}
As illustrated in Fig.~\ref{fig:rdspt} (right)~\cite{cite:Belle:2017}, the result is consistent with the SM expectations. The precision of $R(D^*)$ is 16\%, which is comparable to 9--14\% for the previous measurements with $\lnunu$. With the current statistics, the result excludes $P_\tau(D^*) > +0.5$ at 90\% confidence level. This is the first measurement of $P_\tau(D^*)$ in $\dsttaunu$. 

Figure~\ref{fig:ddsttaunu_summary} illustrates the current situation for the $R(D^{(*)})$ studies summarized by HFAG~\cite{cite:HFAG:2016}. The discrepancy between the world-average $R(D^{(*)})$ and the SM expectations remains at 3.9$\sigma$. This is because the new $R(D^*)$ results from Belle has made the world average closer to the SM but the $R(D^*)$ precision becomes better. The one $R(D)$ and three $R(D^*)$ results from Belle are compatible with the SM expectations within about $2\sigma$ while they tend to be consistently larger than the SM. Our results also agree with the other results by BaBar and LHCb within the current uncertainties. The discrepancy needs to be investigated further at the Belle~II experiment, where 50 times more statistics will be available.
 
\subsection{$\taunu$}

\begin{figure}[b!]
  \centering
  \includegraphics[width=10cm]{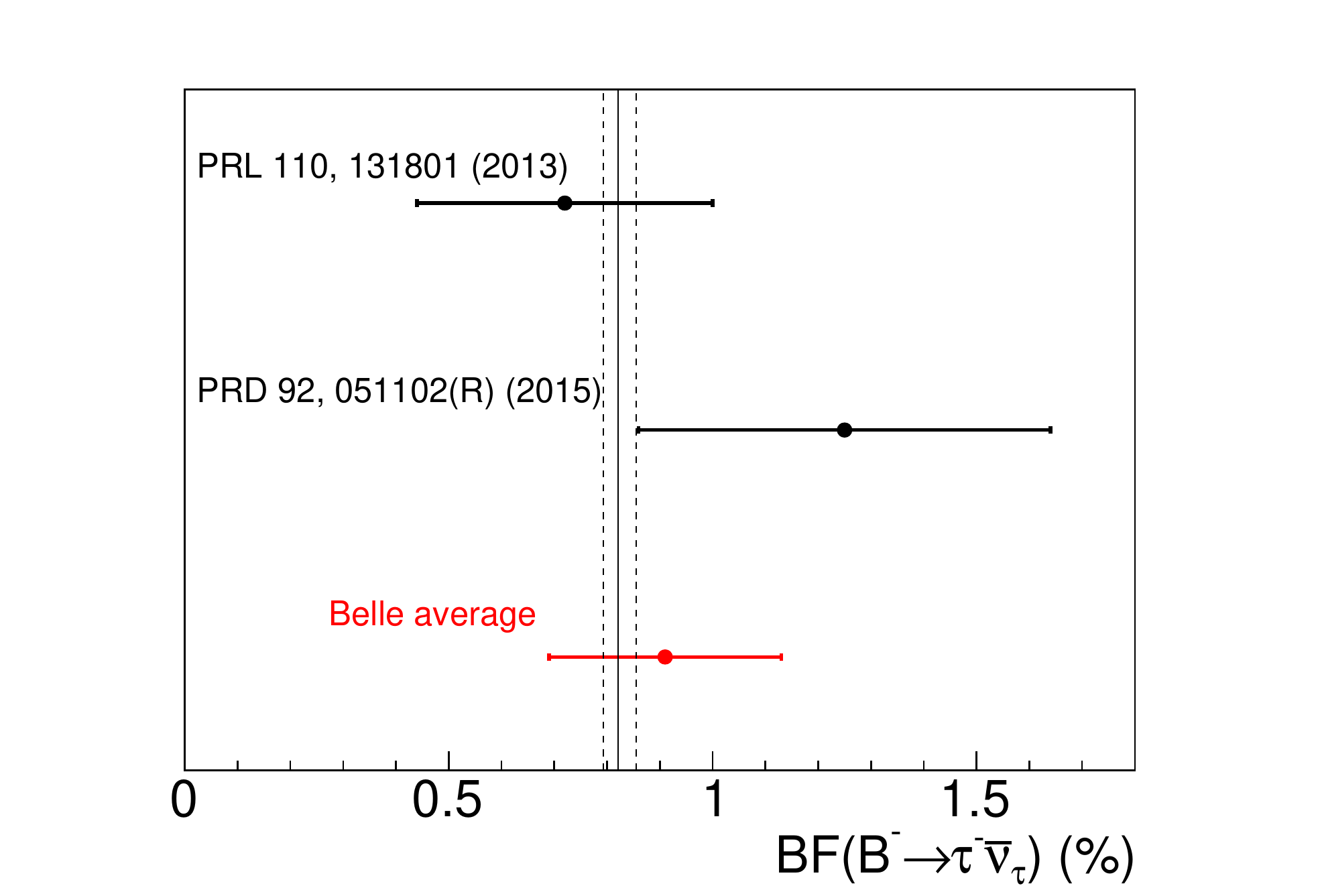}
  \caption{Results from the $\taunu$ measurements at Belle. The vertical line shows the SM expectation with the $\pm 1\sigma$ region.}
  \label{fig:taunu_summary}
\end{figure}

The decay $\taunu$ is one of the purely-leptonic decays of the $B$ meson. Due to the small value of the CKM matrix element $|V_{ub}|$, the branching fraction is suppressed to be $\mathcal{O}(10^{-4})$.

Belle has performed two measurements using the hadronic tagging and the semileptonic tagging~\cite{cite:Belle_taunu:2013,cite:Belle_taunu:2015}. Figure~\ref{fig:taunu_summary} is the comparison of these results with the SM expectation from the preliminary estimation as of ICHEP 2016 by the CKM fitter group~\cite{cite:CKM_fitter}. Their average is consistent with the SM expectation, and the significance is 4.0$\sigma$~\cite{cite:Belle_taunu:2015}.

\subsection{Discussion for the type-II Two-Higgs-Doublet Model}

\begin{figure}[b!]
  \centering
  \includegraphics[width=15cm]{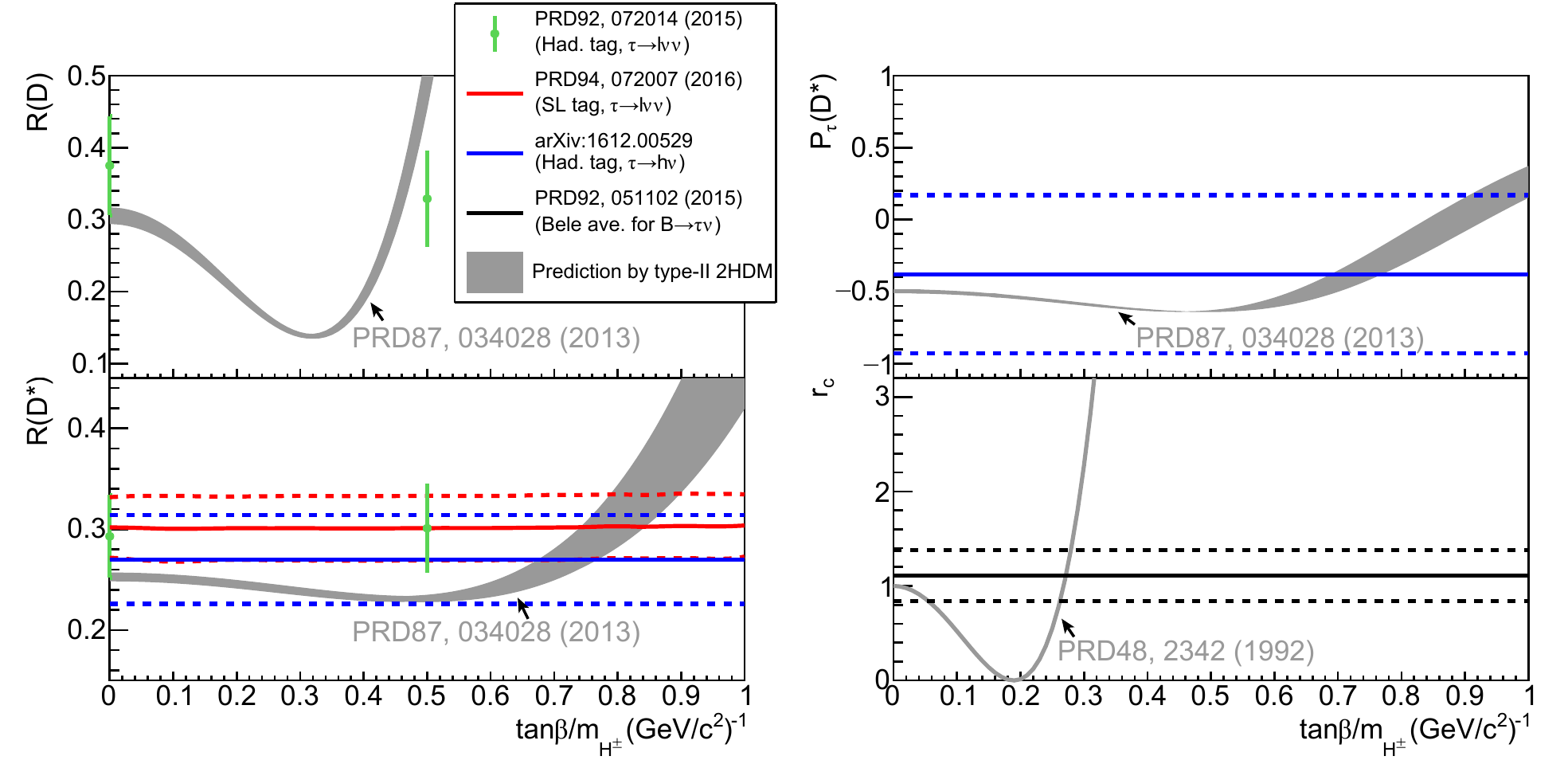}
  \caption{Comparison between the measurements at Belle and the predictions based on the type-II 2HDM for $R(D)$ (top-left), $P_\tau(D^*)$ (top-right), $R(D^*)$ (bottom-left) and $r_c$ (bottom-right). In the $R(D^*)$ and $P_\tau(D^*)$ measurement with the hadronic $\tau$ decays ($\tau^- \rightarrow h^- \nu_\tau$) and the $r_c$ measurement, the efficiency is assumed to be uniform over $\tan \beta / m_{H^\pm}$. The other results include the efficiency correction as a function of $\tan \beta / m_{H^\pm}$.}
  \label{fig:2HDM}
\end{figure}

One of the prominent NP models possibly contributing to the decays $\ddsttaunu$ and $\taunu$ is the Two-Higgs-Doublet Model (2HDM) of type-II~\cite{cite:Gunion:2000}. In this model, the charged Higgs boson $H^\pm$ appears from the additional degrees of freedom in the Higgs doublets and has a large coupling to $b$ and $\tau$. Based on the effective field theory, its contribution is represented by the Lagrangian~\cite{cite:Tanaka:2013,cite:Hou:1992}
\begin{eqnarray}
  \mathcal{L}_{\rm eff} &=& -2 \sqrt{2} G_F V_{ib} \left(\mathcal{O}_{\rm V_1} - m_b m_\tau \frac{\tan^2 \beta}{m_{H^\pm}} \mathcal{O}_{S_1} \right)~(i = u, c),
\end{eqnarray}
where $G_F$, $m_b$ and $m_{H^\pm}$ are the Fermi constant, the masses of the $b$ quark and the charged Higgs, respectively. The parameter $\tan \beta$ denotes the ratio of the vacuum expectation values in the two Higgs doublets. The effective operators $O_{\rm V_1}$ and $O_{S_1}$ corresponds to the SM- and the scalar-type interactions, respectively. See Ref.~\cite{cite:Tanaka:2013} for the explicit definition of these operators. According to this effective Lagrangian, the amplitude of the type-II 2HDM negatively interferes with the SM amplitude.

Belle has measured four observables for $\ddsttaunu$ and $\taunu$: $R(D)$, $R(D^*)$, $P_\tau(D^*)$ and the branching fraction for $\taunu$. Figure~\ref{fig:2HDM} compares our measurements with the predictions from the type-II 2HDM. In this figure, $r_c$ denotes the ratio of the measured or theoretically-expected branching fraction to the SM expectation. At the large $\tan \beta / m_{H^\pm}$ region, $R(D)$ and $r_c$ favors different values of $\tan \beta / m_{H^\pm}$, and this region seems disfavored.

\section{Conclusion}

The decays $\ddsttaunu$ and $\taunu$ are interesting $B$ decays in terms of their sensitivities to NP coupling to $\tau$ leptons, such as the charged Higgs in the type-II 2HDM. In 2016, Belle has shown two new $\dsttaunu$ measurements. One of them is the first application of the semileptonic tagging to the $R(D^*)$ measurement. The second includes the first measurement of $R(D^*)$ using only hadronic $\tau$ decays and the first experimental study of $P_\tau(D^*)$.

We have discussed compatibility of our $\ddsttaunu$ and $\taunu$ measurements with the SM and the type-II 2HDM. All the observables measured by Belle are consistent with but higher than the SM expectations at the 2-$\sigma$ level. For the type-II 2HDM, our results seem to favor the region with small values of $\tan \beta / m_{H^\pm}$.

The world-average $R(D^{(*)})$ including measurements at BaBar and LHCb shows the 3.9-$\sigma$ discrepancy from the SM expectations. With the current precision at Belle, our result is consistent both with the results from these experiments and the SM expectations within about 2$\sigma$. This is an important topic to be further investigated with a high precision at Belle~II. 

\section*{Acknowledgments}

This work was partially supported by JSPS Grant-in-Aid for Scientific Research (S) ``Proving New Physics with Tau-Lepton'' (No.~26220706).

\end{document}